\documentstyle[twocolumn,prb,aps,psfig]{revtex}
\begin{document}
\title{Nonclassical effects and off-diagonal couplings in a model
for FeBr$_2$}
\author{M. Pleimling and W. Selke}
\address{Institut f\"ur Theoretische Physik B, Technische Hochschule,
D--52056 Aachen, Germany}
 
\maketitle
 
\begin{abstract}
Using Monte Carlo techniques, we show that the recently 
experimentally observed transitionlike
phenomena in the transverse spin--ordering close to
the anomalies in the antiferromagnetic
phase of FeBr$_2$ (Petracic et al. [Phys.\ Rev.\ B {\bf 57}, R11051 (1998)])
may result from the quantum nature
of the S=1 spins and an off--diagonal exchange between axial and
planar spin components.
\end{abstract}

\vspace{1cm}

 
Recently, Petracic et al. \cite{pet} presented experimental
evidence for a weakly first--order phase transition in the
antiferromagnetic phase of FeBr$_2$, when applying 
a magnetic field under non--vanishing tilting angle with
respect to the c-axis of the hexagonal crystal, extending previous work
with an axial field. \cite{klee,katsu} They argue that the
measured jumps in the magnetization parallel and perpendicular
to the field are related to the surprisingly sharp peaks near
the anomalies in the specific heat which had been measured
before in axial magnetic fields. \cite{kato}

They suggest that quantum effects in a S=1 anisotropic Heisenberg
magnet and off--diagonal exchange between axial and planar
spin components allowed by the crystal symmetry \cite{mukamel} may
be crucial in explaining the transitionlike phenomena. 

Indeed, our analyses on an appropriate model support their
suggestion. In particular, we considered the following S = 1
Hamiltonian, being believed to describe quite realistically
FeBr$_2$ \cite{mukamel,vettier,her,pouget,selke,pleim}, 
\begin{eqnarray}
{\cal H} & = & -\sum  J_{ij} (\alpha^{-1} S_i^z S_j^z +
 S_i^x S_j^x + S_i^y S_j^y) 
 -\sum D (S_i^z)^2 \nonumber \\
& & -\sum (H_x S_i^x + H_z S_i^z)
 + {\cal H}_{od} \left( J_{xz} \right).
\end{eqnarray}
The first term describes anisotropic, $\alpha < 1$, exchange interactions,
J$_{ij}$, between nearest 
(J$_1$), next-nearest (J$_2$), and third (J$_3$) neighbor spins in
the triangular layers perpendicular to the c--axis and between
spins in adjacent layers connected by equivalent exchange paths (J').
\cite{vettier,her,pouget,selke,pleim}
The second term describes the single--ion anisotropy of
strength D. The third term refers to the tilted field, having 
components both perpendicular (H$_z$) und parallel (H$_x$) to the 
triangular planes. The last term ${\cal H}_{od} \left( J_{xz} \right)$
describes the off--diagonal bilinear couplings between axial (S$_z$)
and planar (S$_x$ and S$_y$) components of neighboring spins in the layers,
with the strength $J_{xz}$.\cite{mukamel} For instance, for neighboring
spins with the same value of $y$, one has
\begin{equation}
{\cal H'}_{od} = - J_{xz} \left( S_i^z S_j^x + S_i^x S_j^z \right) .
\end{equation}
The corresponding terms for other pairs of nearest-neighbor spins follow
from appropriate translations and rotations.\cite{mukamel}

We performed Monte Carlo simulations on that Hamiltonian in
the quasi--classical approximation, i.e. fixing the spin length
to be $\sqrt{2}$, with S$_z$ = $\pm 1, 0$. The exchange 
constants $J_{ij}$ and the anisotropy factor $\alpha$ were
chosen as before \cite{pleim} (based on spin--wave analyses\cite{vettier,pouget}), 
with the tenfold
coordinated antiferromagnetic couplings J' between adjacent
triangular layers, with nearest-neighbor, J$_1$= -16.75 J', and competing
third neighbor, J$_3$ = -0.29 J$_1$, interactions in
the planes perpendicular to the c-axis,
and with an Ising--type anisotropy factor $\alpha =0.78$. 
We then varied the degree of the Ising--like anisotropy due to 
the single ion term, D, the axial, H$_z$, and planar, H$_x$,
components of the external magnetic field, as well as the
strength, J$_{xz}$, of the off--diagonal
exchange between the axial and
planar spin components. 

In an axial field, H$_x$ = 0 and H$_z$ being close to the 
critical field, H$_{z0}$ = $-2$ J' (setting the Boltzmann constant
equal to one), at zero temperature, the peculiar shape of the
temperature dependent specific heat $C(T)$ found
experimentally \cite{kato,pet} can, indeed, be reproduced by tuning
the remaining parameters of the model, as shown in Fig. 1. In particular, the
maximum at the lower temperature shows a pronounced shoulder, associated
with the anomaly in the z--components of the spins \cite{klee,katsu,kato} 
originating from the relatively weak, albeit
dominating, intralayer
couplings as compared to the interlayer couplings as well as from the 
high interlayer coordination number \cite{selke,pleim,voll}. The
superimposed peak shifts towards the critical temperature, T$_c$, and
gets sharper as J$_{xz}$ increases. That peak, at T$_{xy}$, reflects
the disordering of the antiferromagnetic low
temperature state of the xy-- or planar
components of the spins, being aligned ferromagnetically in each
in the triangular planes, with an antiferromagnetic arrangement
between subsequent layers. The corresponding disordering of the
z-- or axial components occurs at T$_c$. The two processes are largely
decoupled due to the quantization of the spins. Note that the
anomaly and the transitionlike peak are not intimately related, and
their positions can be moved relatively to each other by changing, for instance, 
the off--diagonal exchange J$_{xz}$ and the degree of Ising--like
anisotropy D.      
 
Applying a non--axial field, H$_x > 0$ and keeping
H$_z$ fixed, the xy--components of
the spins order in the antiferromagnetic phase, at low temperatures, 
in a spin--flop
state, in which the x--components of the magnetization per layer are 
almost identical in all layers, but the y-components 
of the magnetization in adjacent planes have opposite signs.
Increasing the temperature, the y--component of the magnetization
per layer changes rapidly at the temperature T$_{xy}$,
well below the transition
temperature T$_c$ of the antiferromagnetic phase, see the
Monte Carlo data depicted in Fig. 2. The
change in the y--component may be associated with a
first--order transition, characterized by a jump in the magnetization per
layer, when the off--diagonal exchange J$_{xz}$ exceeds a 
critical value (at about J$_{xz}$ = 18 J'). 
Otherwise, one finds a drastic, but presumably
analytic change. Experimentally \cite{pet}, one is  
at the border of these two scenarios, observing a 'weakly
first--order' transition at T$_{xy}$.--
Note that the experimental findings\cite{pet} are compatible with
the spin--flop state.

In summary, our simulational data show that 
inclusion of the off--diagonal
couplings in the spin components allowed by the
crystal symmetry  
and of the quantum nature of the S=1 spins in the Hamiltonian for
FeBr$_2$ is, indeed, sufficient to describe the transitionlike
features in the specific heat and the magnetization observed 
experimentally recently. The features are related to a rapid change or
transition in
the planar spin components, which, in turn, seems to occur rather
fortuitously closely to the anomaly of the axial spin
components in the antiferromagnetic
phase of FeBr$_2$ discussed extensively before.

\vspace{2cm}

\noindent {\bf Acknowledgement}
 
We thank W. Kleemann for very useful information.

\begin{figure}
\centerline{\psfig{figure=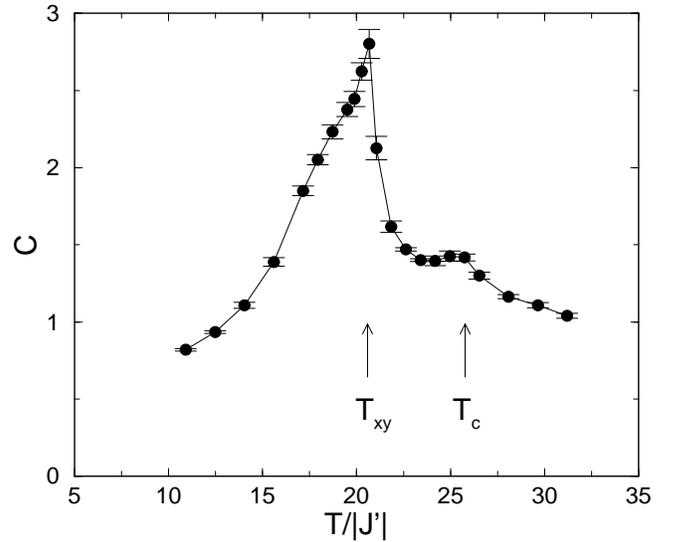,width=8.5cm,angle=270}}
\vspace*{0.2cm}
\caption{Temperature dependent specific heat C as obtained from Monte Carlo
simulations of Hamiltonian (1) with J$_{xz}$= 16.2 J', D= --8.1 J',
and H$_z$= -1.8 J' (= 0.9H$_{z0}$), for systems of 20$^3$ spins.}
\label{fig1} \end{figure}
 
\begin{figure}
\centerline{\psfig{figure=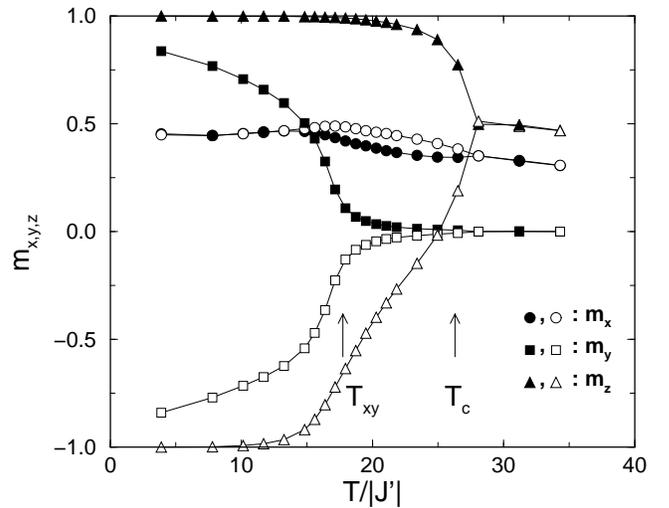,width=8.5cm,angle=270}}
\vspace*{0.2cm}
\caption{Temperature dependence of the x-, y- and z--components
of the magnetization per layer in odd (full symbols) and even
(open symbols) planes, using the 
same parameters as in Fig. 1, with an additional planar
component of the field, H$_x$= 0.75 H$_z$, simulating
systems of 30$^3$ spins.}
\label{fig2}
\end{figure}

\end{document}